\documentclass[preprint,preprintnumbers,amsmath,amssymb]{revtex4}
\usepackage{amssymb,latexsym,amsmath}
\usepackage[activeacute,english]{babel}
\usepackage{fancyhdr}
\usepackage{textcomp}
\usepackage{amsfonts}
\usepackage{graphicx}
\usepackage{subfigure}
\usepackage{graphics}
\usepackage{color}
\usepackage{array}

\begin{document}

\title{Three-loop Correction to the Instanton Density. I. The Quartic Double Well Potential}

\author{M.A.~Escobar-Ruiz$^1$}
\email{mauricio.escobar@nucleares.unam.mx}

\author{E.~Shuryak$^2$}
\email{edward.shuryak@stonybrook.edu}

\author{A.V.~Turbiner$^{1,2}$}
\email{turbiner@nucleares.unam.mx, alexander.turbiner@stonybrook.edu}

\affiliation{$^1$ Instituto de Ciencias Nucleares, Universidad Nacional Aut\'onoma de M\'exico,
Apartado Postal 70-543, 04510 M\'exico, D.F., M\'exico}

\affiliation{$^2$  Department of Physics and Astronomy, Stony Brook University,
Stony Brook, NY 11794-3800, USA}

\date{November 1, 2015}

\begin{abstract}

This paper deals with quantum fluctuations near the classical instanton configuration. Feynman diagrams in the instanton background are used for the calculation of the tunneling amplitude (the instanton density) in the three-loop order for quartic double-well potential. The result for the three-loop contribution coincides in six significant figures with one given long ago by J.~Zinn-Justin. Unlike the two-loop contribution where all involved Feynman integrals are rational numbers, in the three-loop case Feynman diagrams can contain irrational contributions.

\end{abstract}

\maketitle

\section*{Introduction}

\hspace{0.4cm} There is no question that instantons \cite{Polyakov}, Euclidean classical solutions of the
field equations, represent one of the most beautiful phenomena in theoretical physics \cite{VZN}-\cite{Coleman}.
Instantons in non-Abelian gauge theories of the QCD type are important component of the non-perturbative vacuum
structure, in particular they break chiral symmetries and thus significantly contribute to the nucleon (and our)
mass \cite{Schafer:1996wv}. Instantons in supersymmetric gauge theories lead to derivation of the exact beta function \cite{Novikov:1985rd}, and in the ``Seiberg-Witten" $\cal N$=2 case to derivation of the super potential by the exact evaluation of the instanton contributions  to all orders \cite{Nekrasov:2002qd}. The instanton method now has
applications in stochastic settings beyond quantum mechanics or field theories, and even physics -- in chemistry
and biology -- see e.g. discussion of its usage in the problem of protein folding in \cite{Faccioli}.

\hspace{0.4cm} Since the work by A.~Polyakov \cite{Polyakov} the problem of a double well potential (DWP)
has been considered as the simplest quantum mechanical setting illustrating the role of instantons in more
complicated quantum field theories.  In the case of the DWP one can perform certain technical tasks -- like
we do below -- which so far are out of reach in more complicated/realistic settings.

\hspace{0.4cm} Tunneling in quantum mechanical context has been studied extensively using WKB and
other semiclassical means. The  aim of this paper is not to increase accuracy on these quantum-mechanical results,
but rather to develop tools - Feynman diagrams on top of an instanton - which can be used
in the context of many dimensions and especially in Quantum Field Theories (QFTs). Therefore we do $not$ use
anything stemming from the Schr\"odinger equation in this work, in particularly do not use series resulting from
recurrence relations or resurgence relations (in general, conjectured) by several authors.

\hspace{0.4cm} Another reason to study DWP is existing deep connections between the quantum mechanical
instantons -- via Schr\"odinger equation -- with wider mathematical issues, of approximate solutions
to differential equations, defined in terms of certain generalized series. A particular form of an exact
quantization condition was \textit{conjectured} by J.~Zinn-Justin and collaborators (for a review see
\cite{J. Zinn-Justin} and references therein), which links series
around the instantons with the usual perturbative series in the perturbative vacuum.
Unfortunately, no rigorous proof of such a connection exist, and it remains unknown
if it can or cannot be generalized  to the field theory cases we are mainly interested in.
Recently, for the quartic double well and Sine-Gordon potentials Dunne and \"{U}nsal (see \cite{V. Dunne}
and also references therein) have presented more arguments for this connection, which they call
{\it resurgent relation}.

\hspace{0.4cm} In \cite{1-E. Shuryak} the method and key elements (a non-trivial instanton background and
new effective vertices) to calculate the two-loop correction to the tunneling amplitude for the DWP were
established. In particular, the anharmonic oscillator was considered in order to show how to apply Feynman
diagrams technique. In \cite{Olejnik} the Green function in the instanton background was corrected, and
it was attempted to obtain two- and three-loop corrections.  Finally, W\"{o}hler and Shuryak \cite{E. Shuryak}
corrected some errors made in \cite{Olejnik} and reported the exact result for the two-loop correction.

\hspace{0.4cm}The goal of the present paper is to evaluate the three-loop correction to the tunneling
amplitude and compare it with the results  obtained in \cite{J. Zinn-Justin} by a completely
different method, not available in the field theory settings.

\section*{Three-loop correction to the instanton density}

Let us consider the quantum-mechanical problem of a particle of mass $m=1$ in a double well potential
\begin{equation}
V \ = \  \lambda\,{(x^2-\eta^2)}^2 \ .
\end{equation}
The well-known instanton solution $X_{inst}(t) \, = \, \eta\,\tanh(\frac{1}{2}\omega(t-t_c))$, with $\omega^2=8\,\lambda\,\eta^2$, describing the barrier tunneling is the path which possesses the minimal action $S_0=S[X_{inst}(t)]=\frac{\omega^3}{12\,\lambda}$.
Setting $\omega \ =\ 1$, and shifting coordinate to the minimum of the potential one gets the anharmonic oscillator potential in a form $V_{anh} \, =\, \frac{1}{2}x^2-\sqrt{2\,\lambda}\,x^3 + \lambda \, x^4 $
with one (small) dimensionless
parameter $\lambda$.  J.~Zinn-Justin et al \cite{J. Zinn-Justin} use the same potential with $\lambda=g/2$.

The classical action $S_0$ of the instanton solution is therefore large and $\frac{1}{S_0}$ is used
in the expansion.The ground state energy $E_0$ within the zero-instanton sector (pure perturbation theory)
is written in the form
\begin{equation}
E_0 \ = \ \frac{1}{2}\,\sum_{n=0}^\infty \frac{A_n}{S_0^n} \quad , \qquad (A_0=1) \ ,
\label{E0}
\end{equation}
Another series to be discussed is the splitting $ \delta\,E \ = \ E_{first\ excited\ state} - E_{ground \ state}$
related to  the so called instanton density
\footnote{In other words, the energy gap. It was calculated with high accuracy variationally
\cite{Turbiner:2010} and numerically (from thousands to a million of decimals) \cite{Kare}}
in the one-instanton approximation as
\begin{equation}
\delta\,E \ = \ \Delta E\, \sum_{n=0}^\infty \frac{B_n}{S_0^n}  \quad , \quad (B_0=1) \ ,
\label{delE0}
\end{equation}
where $\Delta E = 2 \sqrt{\frac{6\,S_0}{\pi}}\,e^{-S_0}$ is the well-known one-loop semiclassical
result \cite{VZN}. Coefficients $A_{n}$ in the series (\ref{E0}) can be calculated using the ordinary
perturbation theory (see \cite{Bender}) while many coefficients $B_n$ in the expansion (\ref{delE0})
were found by J.~Zinn-Justin, 1981-2005 (see \cite{J. Zinn-Justin} and references therein), obtained via
the so called \textit{exact Bohr-Sommerfeld quantization condition}.

Alternatively, using the Feynman diagrams technique W\"{o}hler and Shuryak  \cite{E. Shuryak} calculated
the two-loop correction $B_1=-71/72$ in agreement with  the result by J.~Zinn-Justin \cite{J. Zinn-Justin}.
Higher order coefficients $B_n$ in (\ref{delE0}) can also be computed in this way.
Since we calculate the energy difference, all Feynman diagrams in the instanton background
(with the instanton-based vertices and the Green's function) need to be accompanied by subtraction of
the same diagrams for the anharmonic oscillator, without the instanton (see \cite{1-E. Shuryak} for details).
For $\frac{1}{\Delta E}\gg \tau \gg 1$ it permits to evaluate the ratio
\[
\frac{\langle  -\eta| e^{-H\,\tau} |  \eta \rangle_{inst}}{\langle  \eta| e^{-H\,\tau} | \eta \rangle_{anh}}
\]
where the matrix elements
$\langle  -\eta| e^{-H\,\tau} |  \eta \rangle_{inst}\ ,\ \langle  \eta| e^{-H\,\tau} |  \eta \rangle_{anh}$
are calculated using the instanton-based and vacuum diagrams,
respectively.

The instanton-based Green's function $G(x,y)$
\begin{equation}
G(x,y) \ = \ G^0(x,y)\bigg[2-xy+\frac{1}{4}|x-y|(11-3xy)+{(x-y)}^2\bigg] +
\frac{3}{8}(1-x^2)(1-y^2)\bigg[{\log}G^0(x,y) -\frac{11}{3}  \bigg] \ ,
\label{GF}
\end{equation}
is expressed in variables $x \,=\, \tanh(\frac{t_1}{2}),\,y \,=\,\tanh(\frac{t_2}{2})\ $,
in which the  familiar oscillator Green function $e^{-|t_1-t_2|}$ of the harmonic oscillator is
\begin{equation}
G^0(x,y) \ = \   \frac{1-|x-y|-x\,y}{1+|x-y|-x\,y} \ ,
\label{GF0}
\end{equation}
In its derivation there were two steps. One was  to find a function which satisfies
the Green function equation, used via two independent solutions and standard Wronskian method.
The second step is related to a zero mode: one can add a term $\phi_0(t_1)\phi_0(t_2)$
with any coefficient and still satisfy the equation. The coefficient is then fixed
from orthogonality to the zero mode, see \cite{Olejnik}.

\begin{figure}[htp]
\begin{center}
\includegraphics[width=4.5in,angle=0]{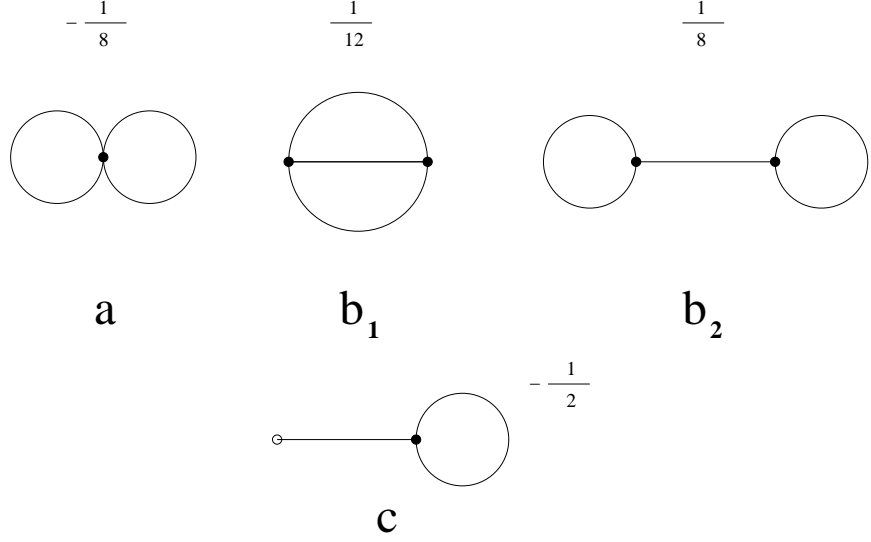}
\caption{Diagrams contributing to the two-loop correction $B_1=a+b_1+b_2+c$.
They enter into the coefficient $B_2$ via the term $B_{2loop}$. For the instanton
field the effective triple and quartic coupling constants (vertices) are $V_3=-\frac{\sqrt{3}}{2}\tanh(t/2)\,S_0^{-1/2}$ and $V_4=\frac{1}{2}\,S_0^{-1}$, respectively, while for the (subtracted) anharmonic oscillator we have $V_3=-\frac{\sqrt{3}}{2}\,S_0^{-1/2}$ and $V_4=\frac{1}{2}\,S_0^{-1}$,
all marked by (filled) bullets.
The tadpole in diagram $c$, which comes from the zero-mode Jacobian rather
than from the action, is effectively represented by the vertex (Jacobian source) $V_{tad}=\frac{\sqrt{3}}{4}\frac{{\tanh}(t/2)}{{\cosh}^2(t/2)}\,S_0^{-1/2}$, marked (unfilled) open bullet.
The signs of contributions and symmetry factors are indicated.}
\label{F1}
\end{center}
\end{figure}

The two-loop coefficient is given by the two-loop diagrams (see Fig. \ref{F1} and \cite{E. Shuryak}),
\[
B_1   =   a+b_1+b_2+c \ ,
\]
where
\begin{equation}
\begin{aligned}
\label{B1}
  a = -\frac{97}{1680}\ , \qquad b_1 = -\frac{53}{1260}\ ,\qquad b_2 = -\frac{39}{560}\ ,
  \qquad c = -\frac{49}{60}\ ,
\end{aligned}
\end{equation}
\bigskip
reflecting the contribution of four Feynman diagrams.

\begin{figure}[htp]
\begin{center}
\includegraphics[width=3.5in,angle=0]{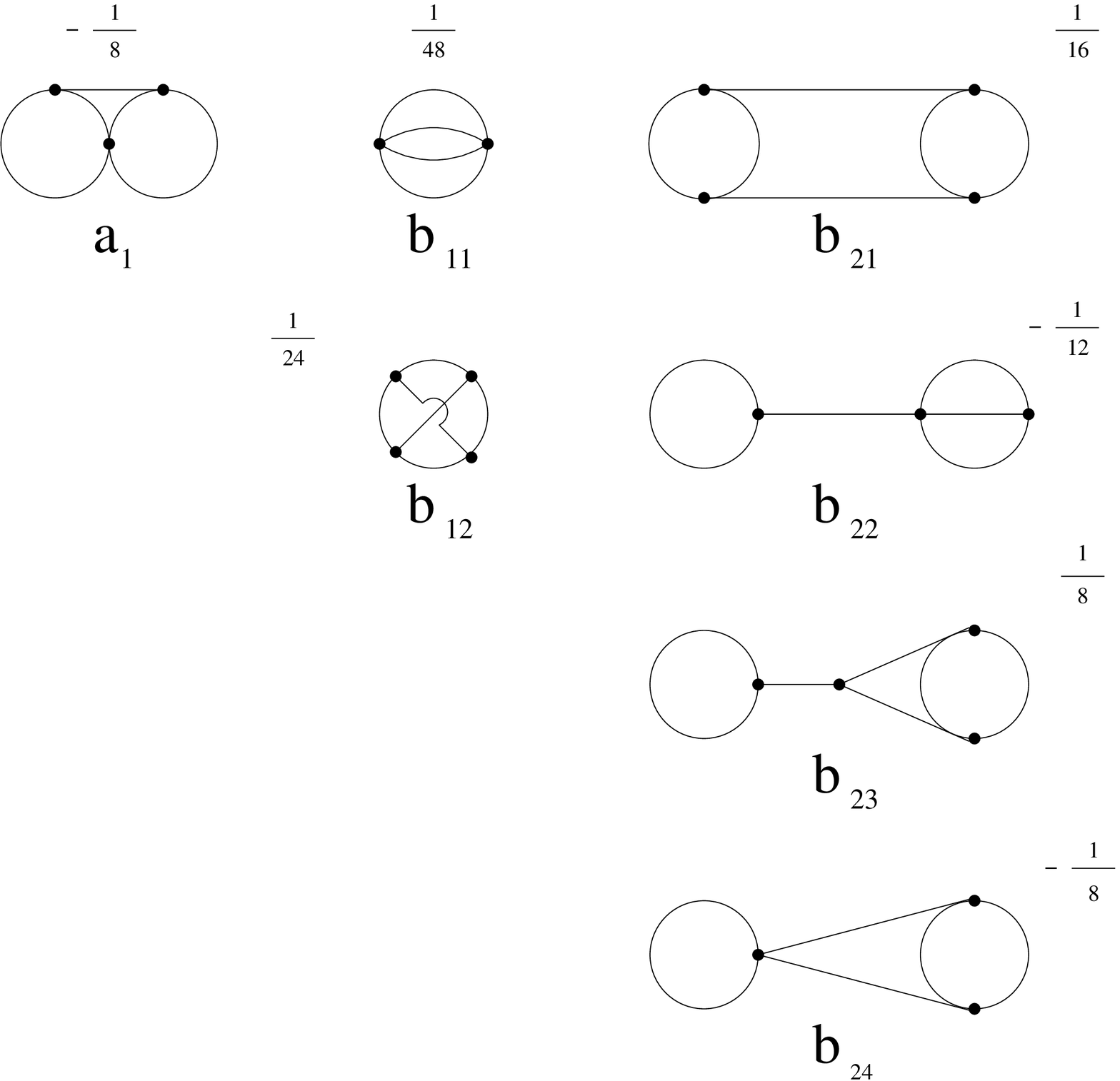}  \vspace{0.8cm}  \\
\includegraphics[width=4.5in,angle=0]{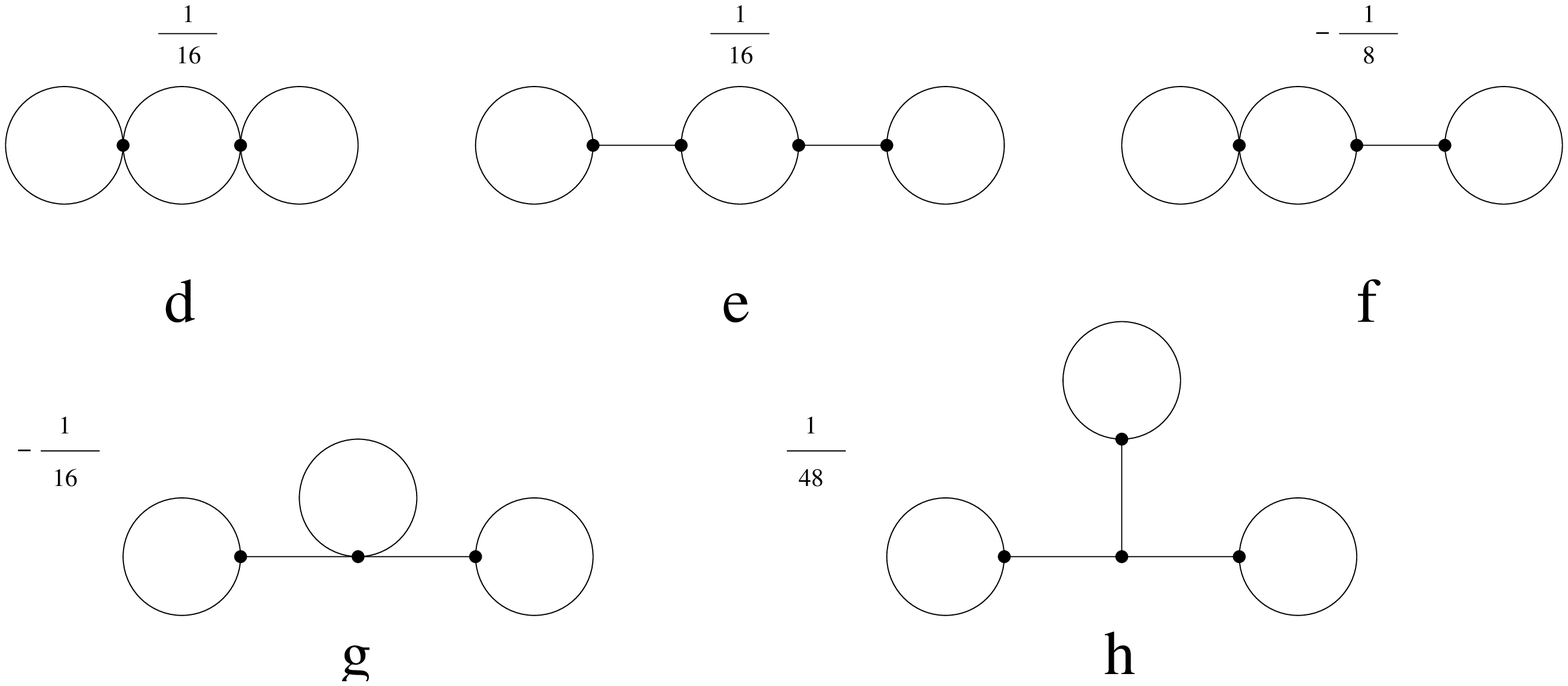}
\caption{Diagrams contributing to the coefficient $B_{2}$. Triple and quartic vertices
$V_3, V_4$ are marked by (filled) bullets.
The signs of contributions and symmetry factors are indicated.}
\label{F2}
\end{center}
\end{figure}

\begin{figure}[htp]
\begin{center}
\includegraphics[width=4.5in,angle=0]{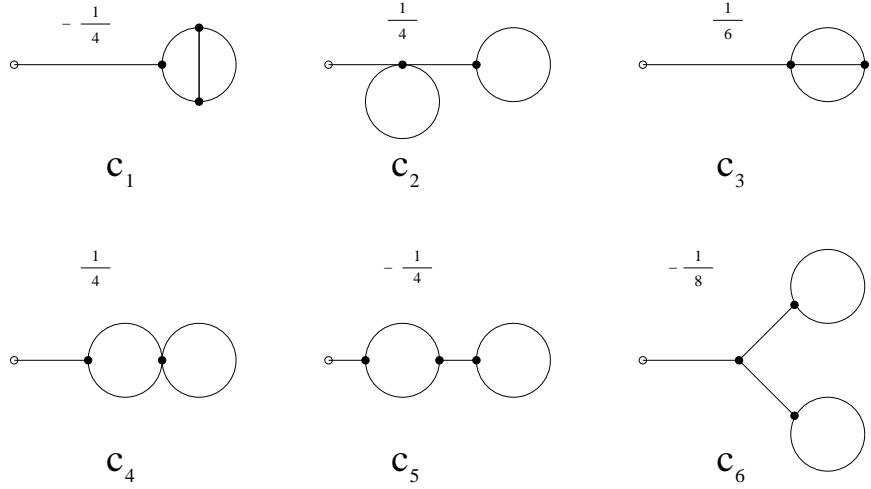}
\caption{Tadpole diagrams contributing to the coefficient $B_{2}$. They come from the Jacobian of the zero mode and have no analogs in the anharmonic oscillator problem. Tadpole vertex $V_{tad}$ (Jacobian source) is marked by (unfilled) open bullet.
The signs of contributions and symmetry factors are indicated.}
\label{F3}
\end{center}
\end{figure}

The three-loop correction $B_2$ (\ref{delE0}) we are interested in is given by the
sum of eighteen 3-loop Feynman diagrams, which we group as follows
(see Figs. \ref{F2} - \ref{F3})
\begin{equation}
\begin{aligned}
 B_{2}  \ = &\  a_1+b_{11}+b_{12}+b_{21}+b_{22}+b_{23}+b_{24}
\\ &  +d+e+f+g+h +c_1+c_2+c_3+c_4+c_5+c_6\ +\ B_{2loop}  \ ,
\label{B2}
\end{aligned}
\end{equation}
\bigskip
complementing by a contribution from two-loop Feynman diagrams
\[
B_{2loop}\ =\ \frac{1}{2}{(a+b_1+b_2)}^2 +(a+b_1+b_2)\,c\ =\ \frac{39589}{259200}\ \ ,
\]
(see (\ref{B1})).
Thus, all Feynman diagrams contributing to (\ref{B2}) are presented in Figs. \ref{F1}-\ref{F3}.
The rules of constructing the integrals for each diagram should be clear from next two
examples. The explicit expression for the Feynman integral $b_{23}$ in Fig. \ref{F2} is
\[
b_{23} =  \frac{9}{8}\int_{-1}^{1}dx\int_{-1}^{1}dy\int_{-1}^{1}dz\int_{-1}^{1}dw\,
\]
\begin{equation}
J(x,y,z,w)\,
\bigg(x\,y\,z\,w\,G_{xx}G_{xy}G_{yz}G_{yw}G_{zw}^2 - G^0_{xx} G^0_{xy} G^0_{yz} G^0_{yw}
{(G^0_{zw})}^2\bigg) \ ,
\end{equation}
while for $c_{4}$ in Fig. \ref{F3} it takes the form
\begin{equation}
c_{4} =  \frac{3}{8}\int_{-1}^{1}dx\int_{-1}^{1}dy\int_{-1}^{1}dz
\frac{x\,y}{(1-y^2)(1-z^2)}\,G_{xy}\,G_{yz}^2\,G_{zz}  \ ,
\end{equation}
here we introduced notations $G_{xy}\equiv G(x,y),\,G^0_{xy}\equiv G^0(x,y)$ and
$J=\frac{1}{(1-x^2)}\frac{1}{(1-y^2)}\frac{1}{(1-z^2)}\frac{1}{(1-w^2)}$. Notice that
the $c$\'{}s diagrams come from the Jacobian of the zero mode and have no analogs
in the anharmonic oscillator (single well potential) problem.

\hspace{0.4cm} For calculation of the symmetry factors for a given Feynman's diagram we use
the Wick's theorem and contractions, see e.g \cite{Palmer-Carrington}. It can be illustrated
by the next two examples.
For diagram $b_{11}$ the four propagators can be rearranged in $4!$ ways and the effect is
duplicated by interchanging the two vertices $t_1,t_2$ giving a symmetry factor of $2\times 4! = 48$.
For the diagram $c_4$ the last propagator, which starts and ends at the same vertex forming loop,
contributes with a factor of two which is also duplicated by rearranging the two inner propagators,
finally, the symmetry factor of $2 \times 2!=4$.

\section*{Results}

The obtained results are summarized in Table \ref{Tab1}. All diagrams are of the form of two-dimensional,
three-dimensional and four-dimensional integrals. In particular, the diagrams $b_{11}$ and $d$
(see Fig. \ref{F2})
\begin{equation}
\begin{aligned}
& b_{11}  \ = \  \frac{1}{48}\int_{-1}^{1}dx\int_{-1}^{1} \frac{dy\,}{(1-x^2)(1-y^2)}
      (\ G^4_{xy} -{(G^0_{xy})}^4  \ )
\\ & d \ \  \ = \  \frac{1}{16}\int_{-1}^{1}dx\int_{-1}^{1}\frac{dy\,}{(1-x^2)(1-y^2)}
   (\ G_{xx}\,G^2_{xy}\,G_{yy}-G^0_{xx}\,{(G^0_{xy})}^2\,G^0_{yy}\ )       \ ,
\end{aligned}
\end{equation}
is given by two-dimensional integrals are the only ones which we are able to calculate analytically
\begin{equation}
\begin{aligned}
& b_{11}  \ = \ -\frac{1842223}{592704000} - \frac{1}{9800}\bigg(367\,\zeta(2)  -
180\,\zeta(3)  - 486\,\zeta(4)  \bigg)\  \equiv b_{11}^{rat} + b_{11}^{irrat} \\
& d \  \  \ = \   \frac{205441}{2469600} + \frac{525}{411600}\zeta(2)\ \equiv d^{rat} + d^{irrat}\ ,
\label{b11}
\end{aligned}
\end{equation}

here $\zeta(n)$ denotes the Riemann zeta function of argument $n$ (see E. Whittaker and G. Watson (1927)).
They contain a rational and an irrational contribution such that
\[
\frac{b_{11}^{irrat}}{b_{11}^{rat}}\approx -4.55 \qquad , \qquad \frac{d^{irrat}}{d^{rat}}\approx 0.025 \ .
\]

It shows that the irrational contribution to $b_{11}$ is dominant with respect to the rational part while
for diagram $d$ the situation is opposite. Other diagrams, see Table I, were evaluated numerically
with an absolute accuracy $\sim10^{-7}$. Surprisingly, almost all of them (15 out of 18 diagrams)
are of the same order $10^{-1}$ as $B_2$ with few of them (diagrams $a_1,\,b_{12},\,b_{21}$)
which are of order $10^{-2}$.

J.~Zinn-Justin (see \cite{J. Zinn-Justin} and references therein) reports a value of
\begin{equation}
B_2^{Zinn-Justin}=-\frac{6299}{10368}\approx -0.6075424\ ,
\label{B2Z}
\end{equation}
while present calculation shows that
\begin{equation}
B_2^{present} \approx \ -0.6075425\ ,
\label{B2O}
\end{equation}
which is in agreement, up to the precision employed in the numerical integration.

Similarly to the two-loop correction $B_1$ the coefficient $B_2$ is negative.
Note also that for  $B_1$ all diagrams are negative while for $B_2$ there are  diagrams of both signs.
For not-so-large barriers ($S_0\sim1$), the two-loop and three-loop corrections are
of the same order of magnitude.

The dominant contribution comes from the sum of the four-vertex diagrams $b_{12},b_{21},b_{23},e,h,c_1,c_5,c_6$
while the three-vertex diagrams $a_1,b_{22},b_{24},f,g,c_2,c_3,c_4$ provides minor contribution, their sum
represents less than $3\%$ of the total correction $B_2$. Interesting
that for both two and three loop cases  the largest
contribution comes from diagrams stemming from the Jacobian, $c$ for $B_1$ and $c_5,c_6$ for $B_2$.
Those diagrams are absent in the perturbative vacuum series, and thus do not have subtractions.

We already noted that some individual three-loop diagrams contain irrational numbers.
Since the J.~Zinn-Justin’s result is a rational number,  then there must be a cancelation of these irrational
contributions in the sum  (\ref{B2}). From (\ref{b11}) we note that the term $(b_{11}^{irrat}+d^{irrat})$
gives a contribution of order $10^{-2}$ to the mentioned sum (\ref{B2}), and therefore the coincidence
$10^{-7}$ between present result (\ref{B2O}) and one of J.~Zinn-Justin (\ref{B2Z}) is an indication that
such a cancelation occurs. Now, we evaluate the coefficients $A_1$, $A_2$ in (\ref{E0}) using Feynman diagrams
(see \cite{Bender}). In order to do it let us consider the anharmonic oscillator potential $V_{anh} \, =\, \frac{1}{2}x^2-\sqrt{2\,\lambda}\,x^3 + \lambda \, x^4 $ and calculate the transition amplitude $\langle x = 0 | e^{-H_{anh}\tau}  | x = 0  \rangle$. All involved Feynman integrals can be evaluated analytically. In the limit
$\tau \rightarrow \infty$ the coefficients of order $S_0^{-1}$ and $S_0^{-2}$ in front of $\tau$ gives us the
value of $A_1$ and $A_2$, respectively. As it was mentioned above the $c$\'{}s diagrams do not exist for the anharmonic oscillator problem. The Feynman integrals in Fig. \ref{F1} give us the value of $A_1$, explicitly they are equal to $$a=\frac{1}{16}\ , \ b_1=-\frac{1}{24}\ ,\ b_2=-\frac{3}{16}\ .$$ The diagrams in Fig. \ref{F2} determine $A_2$ and corresponding values are presented in Table \ref{Tab1}, $b_{11}=-\frac{1}{384}$ and $d=-\frac{1}{64}$.
Straightforward evaluation gives $$A_1=-\frac{1}{3}\ , \ A_2=-\frac{1}{4}\ ,$$ which is in agreement
with the results obtained in standard multiplicative perturbation theory (see \cite{Turbiner:1979-84}).
No irrational numbers appear in the evaluation of $A_1$ and $A_2$. It is worth noting that (see Table
\ref{Tab1}) some Feynman integrals being different give the same contribution,
\[
 f \ =\ g \ =\ \frac{3}{32} \qquad , \qquad b_{22} \ = \ b_{24} \ = \ \frac{1}{24} \  .
\]
In the instanton background the corresponding values of these diagrams do not coincide but are  very close.
\begin{table}[th]
\begin{center}
\setlength{\tabcolsep}{20.0pt}
\begin{tabular}{|c  ||c  | c |}
\hline
Feynman                     &    Instanton        &       Vacuum
\\[1pt]
diagram                     &     $B_2$           &      $A_2$
\\[4pt]
\hline
$a_{1}$           \quad     &   $\, -0.06495185$      &   \quad  $\frac{5}{192}$
\\[3pt]
\hline
$b_{12}$          \quad     &   $\quad 0.02568743$    &     $-\frac{1}{64}$
\\[3pt]
\hline
$b_{21}$          \quad     &   $\quad 0.04964284$    &     $-\frac{11}{384}$
\\[3pt]
\hline
$b_{22}$          \quad     &   $\,-0.13232566$       &    \quad $\frac{1}{24}$
\\[3pt]
\hline
$b_{23}$          \quad     &   $\quad 0.28073249$    &     $-\frac{1}{8}$
\\[3pt]
\hline
$b_{24}$          \quad     &   $\,-0.12711935$       &  \quad   $\frac{1}{24}$
\\[3pt]
\hline
$e$               \quad     &   $\quad 0.39502676$    &     $-\frac{9}{64}$
\\[3pt]
\hline
$f$               \quad     &   $\,-0.35244758$       &  \quad   $\frac{3}{32}$
\\[3pt]
\hline
$g$               \quad     &   $\,-0.39640691$       &  \quad   $\frac{3}{32}$
\\[3pt]
\hline
$h$               \quad     &   $\quad 0.31424977 $   &      $-\frac{3}{32}$
\\[3pt]
\hline
$c_1$             \quad     &   $\,-0.3268200$        &     $-$
\\[3pt]
\hline
$c_2$             \quad     &   $\quad 0.63329511$    &     $-$
\\[3pt]
\hline
$c_3$             \quad     &   $\quad 0.12657122$    &     $-$
\\[3pt]
\hline
$c_4$             \quad     &   $\quad 0.29747446$    &     $-$
\\[3pt]
\hline
$c_5$             \quad     &   $\,-0.77100484$       &     $-$
\\[3pt]
\hline
$c_6$             \quad     &   $\,-0.80821157$       &     $-$
\\[3pt]
\hline
$I_{2D}$           \quad    &   $\quad  0.0963$   &    -$\frac{7}{384}$
\\[3pt]
\hline
$I_{3D}$           \quad    &   $\, -0.0158$      &   \,  $\frac{19}{64}$
\\[3pt]
\hline
$I_{4D}$           \quad    &   $\, -0.8408$      &   -$\frac{155}{384}$
\\[3pt]
\hline
\end{tabular}
\caption{Contribution of each diagram in Fig.\ref{F2} - \ref{F3} for the three-loop corrections $B_2$ (left) and $A_2$ (right) with symmetry factor included. 
We write $B_{2} = (B_{2loop} + I_{2D} +I_{3D} +I_{4D}) $ where $I_{2D},I_{3D},I_{4D}$
denote the sum of two-dimensional, three-dimensional and four-dimensional integrals, respectively.
Similarly, $A_{2}=I_{2D}+I_{3D}+I_{4D}$. The term $B_{2loop}=39589/259200\approx 0.152735$ (see text).}
\label{Tab1}
\end{center}
\end{table}

\section*{Conclusions and Discussion}
In conclusion, we have calculated the tunneling amplitude (level splitting or the instanton density) up to
three-loops using Feynman diagrams for quantum perturbations on top of the instanton. Our result for $B_2$
is found to be in good agreement with the resurgent relation between perturbative and instanton series
suggested by J.~Zinn-Justin (for modern reference see \cite{J. Zinn-Justin}).

\hspace{0.4cm} Let us remind again, that this paper is methodical in nature, and its task was to
develop tools to calculate tunneling phenomena in multidimensional QM or QFT context, in which
any results stemming from the Schr\"odinger equation are not available. We use a quantum mechanical
example as a test of the tools we use: but the tools themselves are expected to work in much wider context.

\hspace{0.4cm} One comment on the results is that the final three-loop answer
has a rational value. However, unlike the evaluation of the two-loop coefficient $B_1$ where all Feynman
diagrams turned out to be rational numbers, in our case of  $B_2$ at least two diagrams contain irrational
parts. What is   the origin of these terms and how they cancel out among themselves are questions left
unanswered above, since several diagrams had resisted our efforts to get the analytic answer, so that
we used numerical multidimensional integration methods, in particular, a dynamical partitioning
\cite{PR:2006}. Perhaps, this can still be improved.

\hspace{0.4cm}  Another intriguing issue is the conjectured relation between the instanton and vacuum series: at the moment we do not understand its origin from the path integral settings. Some diagrams are similar, but expressions quite different and unrelated. New 
(tadpole) diagrams originate from the instanton zero mode Jacobian, and those have no analogues in the vacuum. Surprisingly, they provide the dominant contribution to two-, three-loop corrections $B_1$ (one diagram out of four, see Fig.1) and $B_2$ (six diagrams out of 18, see Fig.3)): $\sim 83\%$ and $\sim 97\%$ \footnote{There is a misprint in this number in the printed version {\it PRD \bf 92}, {\it 025046 (2015)}}, respectively, see Table I. It implies that the sum of vacuum-like diagrams originating from the action is small.

\hspace{0.4cm}  Finally, we note that to our knowledge this is the first three-loop calculation
on a nontrivial background of an instanton. Similar calculations for gauge theories would be certainly
possible and are of obvious interest. One technical issue to be solved is gauge Green function
orthogonal to all (including gauge change) zero modes.

\begin{acknowledgments}
MAER is grateful to J.C.~L\'opez Vieyra for assistance with computer calculations and
for the kind permission to use the code for dynamical partitioning in multidimensional
integration.
This work was supported in part by CONACYT grant {\bf 166189}~(Mexico) for MAER and AVT,
and also by DGAPA grant IN108815-3 (Mexico) for AVT.
The work of ES is supported in part by the U.S. Department of Energy under Contract
No. DE-FG-88ER40388.

\bigskip

\textit{Note added in proof (July, 2015)}:

During the time after the paper was submitted for publication we obtained
a number of new results. We evaluated the contributions of $c,c_5$-like
diagrams, with maximal number of integrations, to the next order
coefficients. The trend continues: those diagrams still contribute
a significant fraction of the total answer, namely $83\%, 127\%, 60\%,
20\%$ of total two-, three-, four-, five-loop $B_1, B_2, B_3, B_4$
contributions, respectively. At the same time, surprisingly, the
absolute values of all these diagrams are rather close. Advance in
numerical multidimensional integrations lead to an increase in
accuracy, the agreement in $B_2$ is now improved to six significant
digits.

\end{acknowledgments}

\section*{Appendix (as of August 9, 2015)}

In this paper for the historical reasons we use the instanton-based Green's function (\ref{GF})
\[
G(x,y) \ = \ G^0(x,y)\bigg[2-xy+\frac{1}{4}|x-y|(11-3xy)+{(x-y)}^2\bigg] +
\frac{3}{8}(1-x^2)(1-y^2)\bigg[{\log}G^0(x,y) -\frac{11}{3}  \bigg] \ ,
\]
expressed in variables $x \,=\, \tanh(\frac{t_1}{2}),\,y \,=\,\tanh(\frac{t_2}{2})\ $,
in which the oscillator Green function
\[
G^0(x,y) \ = \   \frac{1-|x-y|-x\,y}{1+|x-y|-x\,y} \ ,
\]
see (\ref{GF0}).

For the instanton field we have used the effective triple and quartic coupling constants
\[
V_3=-\frac{\sqrt{3}}{2}\tanh(t/2)\,S_0^{-1/2}\ \qquad \ V_4=\frac{1}{2}\,S_0^{-1}\ ,
\]
respectively ($S_0$ is the action evaluated in the instanton solution, see p.3), while for
the (subtracted) anharmonic oscillator we have
\[
V_3=-\frac{\sqrt{3}}{2}\,S_0^{-1/2}\ \qquad \ V_4=\frac{1}{2}\,S_0^{-1}\ .
\]
For tadpole $c$-diagrams the vertex is effectively
represented by
\[
  V_{tad}=\frac{\sqrt{3}}{4}\frac{{\tanh}(t/2)}{{\cosh}^2(t/2)}\,S_0^{-1/2}\ .
\]

%

\hskip 0.6cm The normalization of the Green function (\ref{GF}) was chosen in such a way that in the r.h.s.
of the defining equation contains the coefficient $\frac{1}{2}$ in front of the delta function,
$\frac{1}{2} \delta (x-y)$\,. Vertices $V_3, V_4, V_{tad}$ we have chosen accordingly.

Certainly, the instanton-based Green's function can be derived for the standard normalization
with the coefficient $1$ in front of the delta function $\delta (x-y)$ in the defining equation.
This instanton-based Green's function corresponds to
\[
G(x,y) \rightarrow  G^0(x,y)
\]
\begin{equation}
\bigg[2-xy+\frac{1}{4}|x-y|(11-3xy)+{(x-y)}^2\bigg] +
\frac{3}{8}(1-x^2)(1-y^2)\bigg[{\log}(2\,G^0(x,y)) -\frac{11}{3}  \bigg] \equiv G_n(x,y) \ ,
\label{Gnew}
\end{equation}
where the oscillator Green function changes to
\begin{equation}
G^0(x,y) \rightarrow  \frac{1-|x-y|-x\,y}{2\,(1+|x-y|-x\,y)} \equiv G^0_n(x,y) \ .
\label{G0new}
\end{equation}
Vortices have to be replaced as well
\begin{equation}
V_3 \rightarrow 2\,\sqrt{2}\,V_3 \equiv V_3^n \ ,  \qquad \ V_4 \rightarrow 4\,V_4 \equiv V_4^n\ , \qquad \ V_{tad} \rightarrow \sqrt{2}\,V_{tad} \equiv V_{tad}^n
\label{vertices}
\end{equation}

It can be easily checked that any Feynman integral is invariant under the transformations (\ref{Gnew})-(\ref{vertices}).

\newpage

\clearpage

\end{document}